\documentclass[prb,aps,showpacs]{revtex4}
\usepackage{graphicx,psfig}
\newcommand{\beq}{\begin{equation}}
\newcommand{\eeq}{\end{equation}}

\begin{document}

\title{Strong Coupling Theory for Interacting Lattice Models}

\author{Tudor D. Stanescu and Gabriel Kotliar}
\affiliation{%
Center for Materials Theory, Department of Physics and Astronomy, Rutgers University, Piscataway, New Jersey 08854-8019}
\date{\today}

\begin{abstract}
We develop a strong coupling approach for a general 
lattice problem. We argue that this strong coupling perspective represents the 
natural framework for a generalization of the dynamical mean field theory (DMFT). 
The main result of this analysis is 
twofold: 1) It provides the tools for a unified treatment of any non-local 
contribution to the Hamiltonian. Within our scheme, non-local terms such as 
hopping terms, spin-spin interactions, or non-local Coulomb interactions are 
treated on equal footing. 2) By performing a detailed strong-coupling analysis 
of a generalized lattice problem, we establish the basis for possible  clean 
and systematic extensions beyond DMFT. To this end, we study the problem using 
three different perspectives. First, we develop a 
generalized expansion around the atomic limit in terms of the coupling 
constants for the non-local contributions to the Hamiltonian. By analyzing 
the diagrammatics associated with this expansion, we establish the equations 
for a generalized dynamical mean-field theory (G-DMFT). Second, we 
formulate the theory in terms of a generalized strong coupling version of the 
Baym-Kadanoff functional. Third, following Pairault, S\'en\'echal, and Tremblay\cite{tramb},  we present our scheme in the language of a 
perturbation theory for canonical fermionic and bosonic fields and we 
establish the interpretation of various strong coupling quantities within 
a standard perturbative picture.  
\end{abstract}


\maketitle

\section{Introduction}

Understanding strongly correlated electron systems represents a
major challenge in solid state physics. Dynamical mean-field
theory  (for a review see Ref.\onlinecite{review}) has emerged as a
powerful tool to address this challenge. Within this approach,
the lattice problem for an interacting electron system is mapped
onto a quantum impurity problem with a \textquotedblleft bath\textquotedblright\ determined
self-consistently\cite{dmft1}.  In its simple form, the
construction becomes exact in the limit of infinite lattice
coordination  as shown in Ref. \onlinecite{dmft0}. In addition, several 
generalizations of the approach, such as the extended DMFT\cite{edmft1,edmft2,edmft3,edmft4,edmft5}
 and the DMFT treatment of correlated hopping\cite{schill,shvaik},  
requiring different scalings of the
parameters in the Hamiltonian, have been presented.

More generally,  DMFT  is a tool with strong computational power,
and provides valuable insight into the physics of
strongly-correlated electrons. In the case of simple model
Hamiltonians with purely local interactions, such as the Hubbard
model,  it was able to provide a consistent description for several
non-perturbative properties such as, for example, the Mott-Hubbard metal
insulator transition\cite{review}. Currently, many efforts are
made to extend the DMFT approach in order to make it suitable in
addressing more realistic problems.

The purpose of this work is to develop a strong coupling approach for 
interacting lattice problems. Within this approach, we construct a
systematic
generalization of the DMFT technique capable of describing any
non-local term of a generic lattice Hamiltonian. 
The present formulation has several advantages. First, it unifies various DMFT\
schemes. For example, the treatment of correlated hopping\cite{schill,shvaik}, the
single site DMFT\ of the Hubbard model (see Section III), and the
extended DMFT\cite{edmft1,edmft2,edmft3,edmft4,edmft5} are all 
limiting cases of this unified approach.
Second, besides unifying different DMFT schemes, this formulation is
sufficiently general to be a starting point for expansions around
DMFT. One possibility is described in the Appendix. In addition,
the formulation in terms of Hubbard operators makes it an ideal
framework for a DMFT\ based renormalization group approach \
along the lines of Ref. \onlinecite{rga}.
Third, being based on an expansion around the atomic limit, it
provides new and valuable insight into the diagrammatics
underlying cluster DMFT schemes, which will be a subject of a
subsequent publication. 

\bigskip

The central result of this work is the formulation of a perturbation theory, 
for a general Hamiltonian of the form

\begin{equation}
H=H_{0}+H_{1},   \label{ham1}
\end{equation}%
containing a local term, $H_{0}$, and a non-local one, $H_{1}$. In terms of
Hubbard operators, $X_{i}^{\alpha \beta }$, we can write the two
contributions as
\begin{equation}
H_{0}=\sum_{i}\sum_{\alpha }\lambda _{\alpha }X_{i}^{\alpha \alpha },  \label{ham2}
\end{equation}%
and
\begin{equation}
H_{1}=\sum_{i\neq j}\sum_{\alpha ,\beta ,\alpha ^{\prime },\beta ^{\prime
}}E_{ij}^{\alpha \beta \alpha ^{\prime }\beta ^{\prime }}X_{i}^{\alpha \beta
}X_{j}^{\alpha ^{\prime }\beta ^{\prime }}.    \label{ham3}
\end{equation}%
In these equations $\alpha ,\beta ,\alpha ^{\prime }$ and $\beta ^{\prime }$
represent single-site states. For simplicity, we consider the case of spin $%
\frac{1}{2}$ fermions, so that for each site $i$ there will be four states $%
|\alpha \rangle $: $|0\rangle $ (empty site), $|\uparrow \rangle $ (single
occupied site with spin up), $|\downarrow \rangle $ (single occupied site,
spin down), and $|2\rangle $ (double occupied site). The Hubbard operators $%
X_{i}^{\alpha \beta }$ describe transitions between these states. These
operators have a fermionic character if the occupation numbers of the states
$\alpha $ and $\beta $ differ by one, and bosonic character otherwise. The
algebra of the Hubbard operators is defined by the multiplication rule
\begin{equation}
X_{i}^{\alpha \beta }X_{i}^{\beta ^{\prime }\alpha ^{\prime }}=\delta
_{\beta \beta ^{\prime }}X_{i}^{\alpha \alpha ^{\prime }},    \label{oper1}
\end{equation}%
together with the conserving condition
\begin{equation}
\sum_{\alpha }X_{i}^{\alpha \alpha }=1,
\end{equation}%
and the commutation relation
\begin{equation}
\lbrack X_{i}^{\alpha \beta }X_{j}^{\beta ^{\prime }\alpha ^{\prime }}]_{\pm
}=\delta _{ij}(\delta _{\beta \beta ^{\prime }}X_{i}^{\alpha \alpha ^{\prime
}}\pm \delta _{\alpha \alpha ^{\prime }}X_{i}^{\beta ^{\prime }\beta }).  \label{oper3}
\end{equation}%
In Eq. (\ref{oper3}) the anticommutator is used if both operators are
fermionic and the commutator otherwise. The canonic fermion operators $%
c_{i\sigma }$ can be expressed in terms of Hubbard operators as
\begin{equation}
c_{i\sigma }=X_{i}^{0\sigma }+\sigma X_{i}^{\sigma 2}.
\end{equation}%
The parameters $\lambda _{\alpha }$ in Eq. (\ref{ham2}) represent the
single-site energies and will be determined, in general, by the chemical
potential, $\mu $, the on-site Hubbard interaction, $U$, and the external
fields. The coupling constants $E_{ij}^{\alpha \beta \alpha ^{\prime }\beta
^{\prime }}$ may include contributions coming, for example, from hopping, $%
t_{ij}$, spin-spin interaction, $J_{ij}$, or non-local Coulomb interaction, $%
V_{ij}$. Within the present approach, all these contributions are treated on
equal footing, and we can regard $E_{ij}^{\alpha \beta \alpha ^{\prime
}\beta ^{\prime }}$ as generalized \textquotedblleft
hopping\textquotedblright\ matrix elements.

\bigskip

Our main object of interest is the Green's function of the Hubbard operators.
However, the theory is expressed naturally in terms of a
generalized irreducible two-point cumulant, 
M,  and a dressed hopping,  ${\cal E}$. Using the language of a standard 
field theory (see the Appendix), we interpret M as a generalized self energy 
and ${\cal E}$ as the corresponding Green's function associated with a set of 
auxiliary canonical fermionic and bosonic fields, and show that they have
a clear diagrammatic interpretation in the locator expansion. In addition, 
our formulation contains the average ${\mbox Q}=\langle X\rangle$ of the 
Hubbard operator, which can be 
viewed  as a generalized \textquotedblleft magnetization\textquotedblright, together with a generalized 
effective \textquotedblleft magnetic\textquotedblright\ field, h, represented, in the language of the standard field theory, by the mean value of the auxiliary field. 
Formally, the relationship between these quantities
and the central object of the theory, \ G, is given by the 
equations:
\begin{eqnarray}
G = \langle\langle XX \rangle\rangle - \langle X\rangle\langle X\rangle  = (M^{-1} - E)^{-1},  \nonumber
\end{eqnarray}
and
\begin{eqnarray}
{\cal E} = E(1-ME)^{-1}.        \nonumber      
\end{eqnarray}
where E represents the bare coupling constants.

Next, we introduce the functional
\begin{eqnarray}
\Gamma[{\cal E}, Q] = -\mbox{Tr}~\ln[1-ME] - \mbox{Tr}[M{\cal E}] -\frac{1}{2}QEQ + \Psi[{\cal E}, Q],                \nonumber
\end{eqnarray}
where $\Psi$ represents a generalized Baym-Kadanoff-type functional that 
can be obtained as a sum of all vacuum-to-vacuum skeleton diagrams. Using 
this functional perspective, we can view ${\cal E}$ and M, as well as Q 
and h, as pairs of conjugate quantities and we have
$\delta\Psi / \delta{\cal E} = M$ and $\delta\Psi / \delta Q = h$, where the effective \textquotedblleft magnetic\textquotedblright\ field can be expressed as $h = E Q$.

The single site dynamical mean field theory is  a local 
approximation for \ $\mbox{M} \approx \mbox{M}_{loc}$, 
which can be  obtained as a stationarity condition for the 
functional $\Gamma$. This condition 
translates into  a self-consistent impurity problem defined by the 
statistical operator
\begin{eqnarray}
\hat{\rho}_{imp} = e^{-\beta H_0^{imp}} \hat{T}\exp\left\{
-\int_0^{\beta} X(\tau) \Delta(\tau-\tau^{\prime})
X(\tau^{\prime}) + 
h_{imp}(\tau) X(\tau)\right\},   \nonumber
\end{eqnarray}
where $h_{imp}$ represents an external field and $\Delta$ the  hybridization. 
These quantities are subjected to the  self-consistency conditions:
\begin{eqnarray}
\frac{1}{N}\sum_k ~[M_{loc}^{-1}(i\omega_n) - E(k)]^{-1} =  [M_{loc}^{-1}(i\omega_n) - \Delta(i\omega_n)]^{-1},  \nonumber
\end{eqnarray}
and 
\begin{eqnarray}
h_{imp} + \Delta(0)\langle X \rangle = E\langle X \rangle.   \nonumber
\end{eqnarray}

The strategy that we use in presenting our G-DMFT scheme consists in 
constructing three formally distinct formulations. On 
the one hand, this allows us to establish the equivalence of these 
approaches to the problem of correlated electrons and  to extract a 
generalized unitary picture. On the other hand, these various angles 
 clearly reveal the aspects that are relevant for the 
possible extensions of the theory. 
The natural starting point for our construction is given by a
generalized expansion around the atomic limit. This expansion, in
terms of the coupling constants of the non-local contributions to
the Hamiltonian, is derived in Sec. II. The technique is
characterized by a close formal analogy with the
\textquotedblleft canonical\textquotedblright\ perturbation
expansion in terms of inter-site hopping. Consequently, we will
focus on the new aspects generated by the description of the
problem in terms of Hubbard operators, the main goal being to
obtain the relation between the Green's function for the Hubbard
operators and the irreducible two-point cumulant. Using this
relation, we derive in Sec. III the generalized DMFT equations. 
A simple illustration of the implementation of our scheme is given 
for the Hubbard model. 
In Sec. IV we formulate the theory in terms of a functional of the 
renormalized \textquotedblleft hopping\textquotedblright\. \ The
generalized DMFT equations are shown to be the result of a simple 
local approximation  on a generalized Baym-Kadanoff-type functional. 
An alternative derivation is described in the Appendix. By decoupling 
the non-local term via a Hubbard-Stratonovich transformation,  we 
formulate the theory in terms of a set of canonical fermionic and 
bosonic fields. In this language, the standard treatment based on 
Wick's theorem applies. A possibility of expanding around the DMFT 
solution is also presented.

\bigskip
%$$$$$$$$$$$$$$$$$$$$$$$$$$$$$$$$$$$$$$$$$$$$$$$$$$$$$$$$$$$$$$$$$$

\section{Expansion around the atomic limit}

Expansions around the atomic limit have been powerful tools for studying
both models with localized spins\cite{domb}, and models of interacting
itinerant fermions. Following the pioneering work of Hubbard\cite{hubb1},
Metzner\cite{metz} developed a renormalized series expansion for the
single-band Hubbard model. Generalizing the ``linked-cluster expansion''
ideas\cite{wort}, this approach involves only connected diagrams and
unrestricted lattice sums, and enables one to construct self-consistent
approximations. By analyzing this renormalized expansion, it was shown\cite%
{review} that, in the limit of infinite spatial dimensions, the dynamical
mean-field theory equations for the Hubbard model are recovered. An
extension of this approach to the case of correlated hopping was developed
by Shvaika\cite{shvaik}. The purpose of this section is to obtain a
generalization of the DMFT equations that describe the physics of a generic
lattice model for interacting fermions, starting from a renormalized
strong-coupling expansion around the atomic limit\cite{pank}. Within this scheme, all
the non-local contributions to the Hamiltonian, for example hopping terms,
spin-spin interaction terms, or non-local Coulomb interaction contributions,
are treated on equal footing.

Let us consider a system described by the Hamiltonian (\ref{ham1}).
The first step in our derivation is to write an expansion for the
grand-canonical potential and the Green`s functions for the $X$-operators in
terms of ``hopping'' matrix elements $E_{ij}^{\alpha \beta
\alpha^{\prime}\beta^{\prime}}$ and bare cumulants. Starting with the
statistical operator
\begin{equation}
\hat{\rho} \equiv e^{-\beta H_0} \hat{\sigma}(\beta) = e^{-\beta H_0} \hat{T}
\exp\left\{-\int_0^{\beta} d\tau H_1(\tau)\right\},  \label{sigm}
\end{equation}
we can write the grand-canonical potential as
\begin{equation}
\Omega = \Omega_0 - \frac{1}{\beta} \ln \langle\hat{\sigma}(\beta)\rangle_0 ,
\label{omeg}
\end{equation}
where $\Omega_0 = -\frac{1}{\beta} \ln Tr(e^{-\beta H_0})$ is the grand-canonical potential in the atomic limit, $\beta = 1/k_BT$
is the inverse temperature, and the unperturbed ensemble average is given by
\begin{equation}
\langle ~.~.~.~\rangle_0 = Tr\{\exp(-\beta H_0) ~.~.~.~\}/Tr\{\exp(-\beta
H_0)\}.
\end{equation}
Expanding the exponential in Eq. (\ref{sigm}) we obtain the nth order
contribution to $\langle\hat{\sigma}(\beta)\rangle_0$:
\begin{eqnarray}  \label{expan}
\langle\hat{\sigma}(\beta)\rangle_0^{(n)} &=& \frac{(-1)^n}{n!}%
\int_0^{\beta} d\tau_1~.~.~.~\int_0^{\beta} d\tau_n \langle\hat{T}%
\{H_1(\tau_1)~.~.~.~H_n(\tau_n)\}\rangle_0  \nonumber \\
&=& \frac{(-1)^n}{n!} \sum_{\{i_k, j_k\}}
\sum_{\{\alpha_k,\beta_k,\alpha_k^{\prime},\beta_k^{\prime}\}}
E_{i_1j_1}^{\alpha_1 \beta_1
\alpha_1^{\prime}\beta_1^{\prime}}~.~.~.~E_{i_nj_n}^{\alpha_n \beta_n
\alpha_n^{\prime}\beta_n^{\prime}}\int_0^{\beta} d\tau_1~.~.~.~d\tau_n \\
&\times&\langle\hat{T}\{X_{i_1}^{\alpha_1
\beta_1}(\tau_1)X_{j_1}^{\alpha_1^{\prime}\beta_1^{\prime}}(\tau_1) ~.~.~.~
X_{i_n}^{\alpha_n
\beta_n}(\tau_n)X_{j_n}^{\alpha_n^{\prime}\beta_n^{\prime}}(\tau_n)\}%
\rangle_0,  \nonumber
\end{eqnarray}
where $\hat{T}$ represents the imaginary-time ordering operator, and $%
X_{i}^{\alpha \beta}(\tau)$ are the Hubbard operators in the interaction
representation,
\begin{equation}
X_{i}^{\alpha \beta}(\tau) = e^{H_0\tau}X_{i}^{\alpha \beta}e^{-H_0\tau}.
\end{equation}
Using the multiplication properties (\ref{oper1}) of the Hubbard operators,
it is straightforward to obtain the explicit imaginary time dependence,
\begin{equation}
X_{i}^{\alpha \beta}(\tau) =
e^{-(\lambda_{\alpha}-\lambda_{\beta})\tau}X_{i}^{\alpha \beta}.
\label{xtau}
\end{equation}
Because $H_0$ is a sum of local operators, the ensemble average in (\ref%
{expan}) factorizes into independent local averages that can be evaluated
using the algebra of the X-operators. However, the explicit calculation of $%
\Omega$ is cumbersome, as the summation over the site variables $i_k$ is
restricted. To overcome this problem, following Metzner\cite{metz}, we
introduce the bare cumulant defined by
\begin{equation}
C^n_i(\alpha_1, \beta_1,~.~.~.~,\alpha_n, \beta_n; \tau_1,~.~.~.~,\tau_n ) =
\frac{\delta^n W}{\delta\eta_{\alpha_1
\beta_1}(\tau_1)~.~.~.~\delta\eta_{\alpha_n \beta_n}(\tau_n)}\Bigg|_{\{\eta
= 0\}},  \label{cumul0}
\end{equation}
where the generating functional is
\begin{equation}
W[\{\eta\}] = \ln\Bigg\langle \hat{T}\exp\left\{ - \sum_{\alpha \beta}
\int_0^{\beta} d\tau ~\eta_{\alpha \beta}(\tau) X^{\alpha
\beta}(\tau)\right\}\Bigg\rangle_0.  \label{wfunc}
\end{equation}
The fields $\eta_{\alpha \beta}$ are either complex numbers or Grassmann
variables, depending on the nature (bosonic or fermionic) of the
corresponding operators $X^{\alpha \beta}$. In the Eqs. (\ref{cumul0}) and (%
\ref{wfunc}) the site index, $i$, for the $\eta$ field and the Hubbard
operators, was omitted for simplicity.

\begin{figure}[tbp]
\begin{center}
\includegraphics[width=0.5\textwidth]{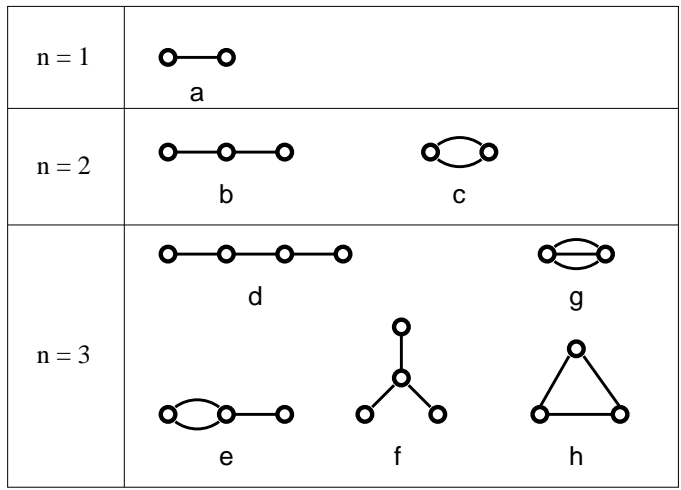}
\end{center}
\caption{Diagrams contributing to the expansion of the grand canonical
potential in terms of generalized ``hopping'' matrix elements (full lines)
and bare cumulants (small circles). The contributions from diagrams a, b, d,
e, f, and g are non-zero only if $H_1$ contains terms with bosonic 
X-operators having non-zero average values (Q-operators).}
\label{FIG1}
\end{figure}

Using the bare cumulants allows one to express the grand-canonical potential
as an expansion containing unrestricted sums. Each term will be a product of
``hopping'' matrix elements, $E_{ij}^{\alpha \beta
\alpha^{\prime}\beta^{\prime}}$, and local cumulants, and can be represented
diagrammatically. Further, using linked-cluster type arguments\cite{wort}
one can show that only the connected diagrams contribute. The basic
diagrammatic rules are similar to those given by Metzner\cite{metz} for the
Hubbard model. However, the internal lines corresponding to ``hopping''
matrix elements will carry two extra pairs of indices $(\alpha\beta)$ and $%
(\alpha^{\prime}\beta^{\prime})$ (one for each end of the line) representing
the single-site states. Also, the nth order cumulant, $C_i^{(n)}$,
represented by a n-valent point vertex, will be attached to n lines, with
the corresponding labels $(\alpha_k \beta_k)$. If we split the states
associated with a cumulant in two subsets, $(\alpha_1,~.~.~.~,\alpha_n)$ and
$(\beta_1,~.~.~.~,\beta_n)$, the multiplication rule (\ref{oper1}) requires
that each single-particle state ($|0\rangle$, $|\sigma\rangle$, or $|2\rangle
$) be found the same number of times in each of the subsets. Also, in
contrast with the simple case of an expansion in the hopping matrix, $t_{ij}$
, in general, cumulants of order n with n odd may be non-zero. This is the 
case if $H_1$ contains terms with bosonic X-operators having a non-zero  
average value. 
Such terms may occur, for example, if a non-local Coulomb interaction
$V_{ij}n_in_j$ is considered. To summarize these observations, we present in 
Fig. \ref{FIG1} the diagrams yielding the leading contributions to $\Omega$.

\begin{figure}[tbp]
\begin{center}
\includegraphics[width=0.5\textwidth]{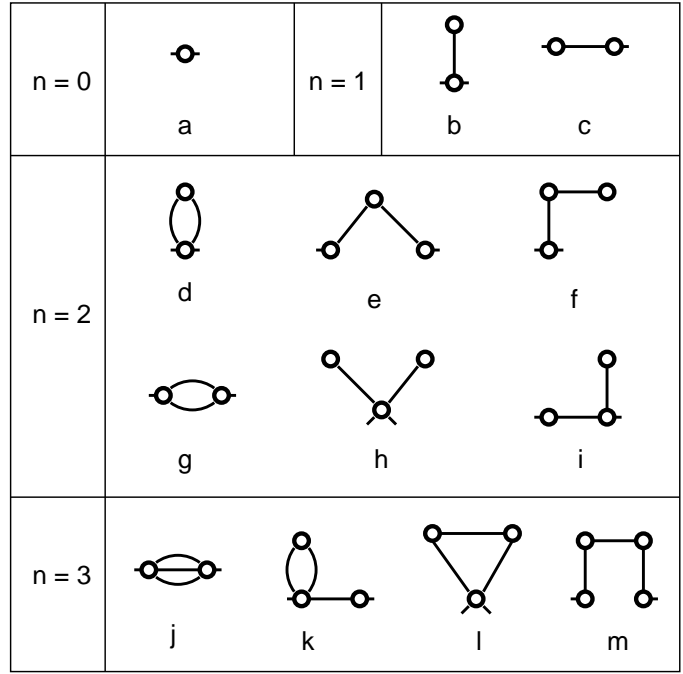}
\end{center}
\caption{Diagrams representing the leading contributions to the expansion of
the Green's function around the atomic limit. The cumulants characterized by
fixed site and state labels have small legs attached to the corresponding 
circle. For $n=3$ only diagrams containing cumulants with an even number of
legs are shown.}
\label{FIG2}
\end{figure}

Our next task is to write an expansion for the Green's functions of the  
Hubbard operators. We define
\begin{equation}
\bar{G}_{\alpha \beta
~\alpha^{\prime}\beta^{\prime}}(i,j;\tau,\tau^{\prime}) = -Tr\{\hat{\rho}~%
\hat{T}[X_i^{\alpha\beta}(\tau)X_j^{\alpha\beta}(\tau^{\prime})]\},
\label{green1}
\end{equation}
where the statistical operator $\hat{\rho}$ is given by Eq. (\ref{sigm}),
and the Hubbard operators are in the Heisenberg representation,
\begin{equation}
X_{i}^{\alpha \beta}(\tau) = e^{H\tau}X_{i}^{\alpha \beta}e^{-H\tau}.
\end{equation}
The perturbation expansion for the Green's functions is similar to the
expansion for the grand-canonical potential. The consequence of the presence
of two extra X-operators in (\ref{green1}) is that the corresponding
diagrams will be rooted, i.e. one (for $i=j$) or two of the vertices will
have fixed site indices and some fixed single-site state indices. The
general rules are analogous to those for the Hubbard model\cite{metz}, with
the difference that extra indices (associated with the single-site states)
are carried by the ``hopping'' matrix elements and the bare cumulants. 
A special case occurs when bosonic X-operators with non-zero average values, 
$\langle X^{\alpha\beta} \rangle = Q^{\alpha\beta}$, which we will call Q-operators, are present in $H_1$. 
 In this
case, as mentioned before, cumulants of order n with n odd may be non-zero,
leading to disconnected rooted diagrams in the expansion of 
$\bar{G}_{\alpha\beta \alpha^{\prime} \beta^{\prime}}$. By simply regrouping the contributions 
from
disconnected diagrams we can write:
\begin{equation}
\bar{G}_{\alpha \beta ~\alpha^{\prime} \beta^{\prime}}(i,j;\tau,\tau^{\prime}) = G_{\alpha \beta ~\alpha^{\prime} \beta^{\prime}}(i,j;\tau,\tau^{\prime}) - \langle
X_i^{\alpha\beta}\rangle \langle X_i^{\alpha^{\prime} \beta^{\prime}}\rangle,  \label{green2}
\end{equation}
where $\langle X_i^{\alpha\beta}\rangle$ represents the average of the 
operator $X_i^{\alpha\beta}$ with respect to the full Hamiltonian, and  
$G_{\alpha\beta \alpha^{\prime} \beta^{\prime}}$ has a diagrammatic representation containing 
only connected diagrams. In the rest of this work we will always use the 
``connected'' Green's functions defined by (\ref{green2}). Obviously, if $H_1$
does not contain Q-operators,
$G$ and $\bar{G}$ coincide, as the averages in (\ref
{green2}) vanish. To summarize, we present in Fig. \ref{FIG2} some leading
contributions to $G$. 

At this point, a brief discussion about the Fourier transformation of the 
Green's functions and of the cumulants is required. Due to the fact that the
Hubbard operators have either bosonic-like or fermionic-like characters, the
Fourier transformed Green's functions, $G_{\alpha \beta
~\alpha^{\prime}\beta^{\prime}}(i,j;i\omega_n)$, will have odd Matsubara
frequency, if both X-operators are fermionic, or even Matsubara frequency
otherwise. On the other hand, the cumulants may depend on both even and odd
Matsubara frequencies. Recalling the definition, Eqs. (\ref{cumul0}-\ref%
{wfunc}), a pair of indices $\alpha_k\beta_k$ associated with a
fermionic-like Hubbard operator will induce a dependence on an odd frequency
$\omega_k$, while an even frequency will be associated with a pair with
bosonic character. A line connecting two vertices, and yielding the factor $%
E_{ij}^{\alpha \beta ~\alpha^{\prime}\beta^{\prime}}$, will carry the
frequency corresponding to the nature of the two pairs $(\alpha \beta)$ and $%
(\alpha^{\prime}\beta^{\prime})$. This nature is always the same as a result
of the fact that the Hamiltonian conserves the number of particles. In the
rest of this work we will not use different notations for odd and even
frequencies, but we should keep in mind this aspect of the problem.

\begin{figure}[tbp]
\begin{center}
\includegraphics[width=0.7\textwidth]{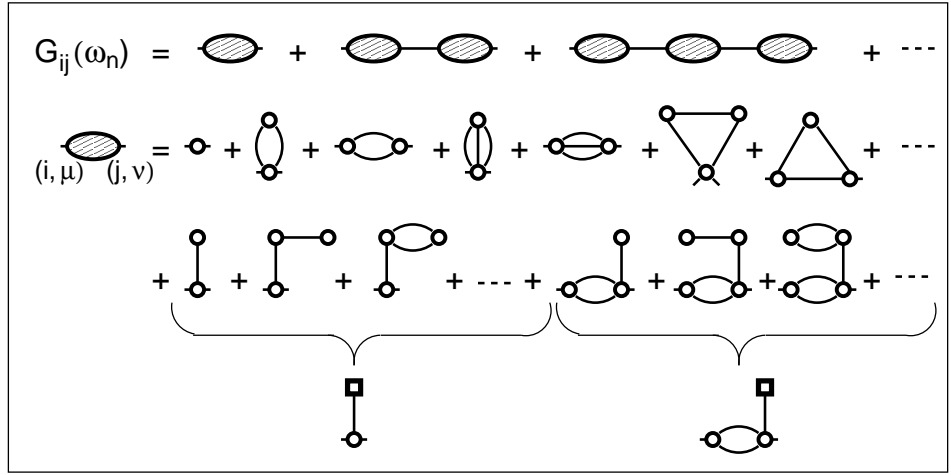}
\end{center}
\caption{Diagrammatic representation of the renormalized expansion of the
Green's function and some leading contributions to the irreducible two-point
cumulant $M_{ij}^{\protect\mu \protect\nu}(i\protect\omega_n)$. The
simplified notation $(\protect\alpha \protect\beta) = \protect\mu$, $(%
\protect\alpha^{\prime}\protect\beta^{\prime}) = \protect\nu$ was used. The
diagrams containing one-particle reducible decorations of bare cumulants can
be summed and replaced by diagrams with decorations equal to $\sum_{j,
\protect\nu} E_{ij}^{\protect\mu~\protect\nu} \langle X_j^{\protect\nu} \rangle$. The average $%
\langle X_j^{\protect\nu} \rangle$ is represented
diagrammatically by a square.}
\label{FIG3}
\end{figure}

The expansion that we derived is rather complicated and it may be of little
help if used directly. However, we know that in order to obtain
self-consistent approximations and, in particular, DMFT-type equations, a
renormalized expansion is necessary. The key in obtaining a renormalized
expansion is to introduce the notion of irreducibility with respect to one
line (representing $E_{ij}^{\alpha \beta ~\alpha^{\prime}\beta^{\prime}}$).
We define the irreducible two-point cumulant, $M_{ij}^{\alpha \beta
~\alpha^{\prime}\beta^{\prime}}(i\omega_n)$, as the sum of all connected
diagrams having two external legs (labeled by $\alpha \beta$ and $%
\alpha^{\prime}\beta^{\prime}$, respectively), with the property that the
two legs cannot be separated by cutting one line. This definition differs
slightly from the standard definition\cite{metz,review}, which requires that
the diagram be not disconnected by cutting a single line. The difference
comes from the existence of cumulants with odd number of legs. If $H_1$ does
not contain Q-operators, the standard diagrammatics, along with the
usual definition of irreducibility are valid. Some typical, low order
contributions to the irreducible cumulant $M_{ij}^{\alpha \beta
~\alpha^{\prime}\beta^{\prime}}(i\omega_n)$ are shown in Fig.\ref{FIG3}.
Notice that we can classify the diagrams into classes, each class containing
one ``basic'' diagram that is one-particle irreducible with respect to all
``hopping'' lines plus diagrams derived from it by decorating the bare
cumulants with reducible contributions. It is useful to perform partial
summations of the diagrams in each class. The resulting diagrams will have
all the reducible decorations replaced by factors equal to $\sum_{j, \alpha^{\prime}, \beta^{\prime}}
E_{ij}^{\alpha\beta~\alpha^{\prime}\beta^{\prime}} \langle X_j^{\alpha^{\prime}\beta^{\prime}} \rangle$.

We can now write a renormalized expansion for the Green's functions in terms
of irreducible cumulants and ``hopping'' matrix elements. The diagrammatic
representation of this expansion is shown in Fig.\ref{FIG3}. Analytically,
the corresponding equation will be:
\begin{equation}
G_{ij}^{\mu\nu} = M_{ij}^{\mu\nu} + \sum_{k,l}\sum_{\rho,\pi}~
M_{ik}^{\mu\rho}E_{kl}^{\rho\pi}G_{lj}^{\pi\nu},  \label{lark1}
\end{equation}
where, to reduce the number of indices, we introduced a simplified notation
for a pair of single-site states, $(\alpha \beta) = \mu$. It is also
convenient to regard a quantity $A^{\mu\nu}$ as an element of a matrix $%
\mathbf{A}$. Using this matrix notation, and Fourier transforming the
spatial dependence, one can write the relation between the irreducible
cumulant and the X-operator Green's function,
\begin{equation}
\mathbf{G}(\mathbf{k}, i\omega_n) = \left[\mathbf{M}^{-1}(\mathbf{k},
i\omega_n) - \mathbf{E}(\mathbf{k})\right]^{-1},  \label{lark2}
\end{equation}
where $\mathbf{E}(\mathbf{k})$ is the Fourier transform of the coupling
constant matrix. What exactly are the Green's function included in the
matrix $\mathbf{G}$? Obviously, those that are coupled by the generalized
``hopping'' $E_{ij}^{\mu\nu}$. Specifically, if $\{\mu_n\}_{n={1,...,P}}$
and $\{\nu_n\}_{n={1,...,P}}$ are the sets of single-site pairs that label
the coupling constants in $H_1$, each pair appearing only once in the
corresponding set, the matrices will have dimension P and the matrix
elements will be labeled $G^{\nu_n \mu_n}$, $M^{\nu_n \mu_n}$, and $E^{\mu_n
\nu_n}$, respectively. A particular example of this construction will be
presented in the next section for the Hubbard model.

Equations (\ref{lark1}) and (\ref{lark2}) represent the main result of this
section. They are valid in arbitrary dimensions and for any Hamiltonian that
has the form given by Eqs. (\ref{ham1}-\ref{ham3}) when written in terms of
Hubbard operators.

%Section II Section II Section II Section II Section II Section II Section II

\section{Generalized DMFT Equations} 

In the standard dynamical mean-field theory\cite{review}, the 
single-particle self energy for the electron
Green's function is local in the limit of infinite spatial dimensions. 
However, for  a general lattice
Hamiltonian containing fermionic X-operators, 
this property is violated, as shown by Schiller\cite{schill} in
the case of correlated hopping. 
Nevertheless, one can overcome this problem by adopting a strong-coupling 
perspective\cite{shvaik} and regarding the two-point irreducible cumulant, 
rather than the self-energy, as the the quantity in which the theory should 
be formulated.  If $H_1$ contains only fermionic X-operators, 
and assuming that the coupling constants in $H_1$ are properly
scaled, to keep the energy (per site) finite, one can show\cite{metz,review}  
that the irreducible cumulants become local,
\begin{equation}
\mathbf{M}_{ij}(i\omega_n) = \delta_{ij}\mathbf{M}_0(i\omega_n).  \label{mm0}
\end{equation}
Consequently,  the natural quantity in which such a theory should be
formulated is the irreducible cumulant, rather than the self-energy.
This becomes immediately transparent using
Eqs. (\ref{lark2}) and (\ref{mm0}), and the fact that the electron Green's
function is given by the sum
\begin{equation}
\mathcal{G}_{\sigma}(\mathbf{k}, i\omega_n) = G^{0 \sigma ~ \sigma 0}(%
\mathbf{k}, i\omega_n) + G^{\bar{\sigma} 2 ~ 2 \bar{\sigma}}(\mathbf{k},
i\omega_n) + \sigma[G^{0 \sigma ~ 2 \bar{\sigma}}(\mathbf{k}, i\omega_n) +
G^{\bar{\sigma} 2 ~ \sigma 0}(\mathbf{k}, i\omega_n)].  \label{elgreen}
\end{equation}
Although $\mathbf{M}_0$ is k-independent, $\mathcal{G}$ will have, in
general, a complicated dependence on the wave-number, and the corresponding
self-energy will be non-local, breaking the standard DMFT scheme. 
Formally, a self-energy matrix can be introduced by defining
\begin{equation}
\mathbf{\Sigma}(z) = (z+\mu)\mathbf{I} - \mathbf{E} - \mathbf{G}^{-1}(z) =
(z+\mu)\mathbf{I} - \mathbf{M}_0^{-1}(z),
\end{equation}
with $\mathbf{I}$ being the identity matrix. However, in general, this
quantity does not have the desired analytic properties\cite{shvaik} and may
diverge as $|z| \rightarrow \infty$.

If we go further and generalize the model to include both fermionic and 
bosonic degrees of freedom, the picture presented above fails, as one 
cannot scale the coupling constants in such a way that one has both finite energy 
(per site) and local irreducible cumulants in the limit of infinite 
dimensions\cite{pkm}. In contrast, this work assumes a different perspective: rather than viewing DMFT as an approximation that becomes exact in the limit of infinite coordination, we see it as the first order approximation in a hierarchy of (increasing size) cluster approximations. In this picture, the exact limit is obtained for an infinite-large cluster. Consequently, our G-DMFT scheme will be defined by the condition that the two-point irreducible cumulant be local.
Unlike the case of extended DMFT, we have no small parameter (other than the inverse of the cluster size) justifying this approach.
 Obviously, if the model contains only fermionic degrees of freedom, the two perspectives described above are equivalent.   
 
Based on the locality of $\mathbf{M}$, we can approach the central step in a
DMFT-type scheme, namely the self-consistent mapping of the lattice problem
onto an effective impurity problem. Let us consider the single-impurity
problem with the statistical operator
\begin{equation}
\hat{\rho}_{imp} = e^{-\beta H_0^{imp}} \hat{T}\exp\left\{
-\int_0^{\beta}d\tau \left[\int_0^{\beta}d\tau^{\prime}~ \sum_{\alpha \beta
~ \alpha^{\prime}\beta^{\prime}} X^{\alpha \beta}(\tau) \Delta^{\alpha \beta
~ \alpha^{\prime}\beta^{\prime}}(\tau-\tau^{\prime})
X^{\alpha^{\prime}\beta^{\prime}}(\tau^{\prime}) + \sum_{\alpha, \beta \in \mathcal{B}_Q}
h_{imp}^{\alpha\beta}(\tau) X^{\alpha \beta}(\tau)\right]\right\},  \label{impur}
\end{equation}
where $h_{imp}(\tau)$ is an external field that couples to the
Q-operators, $\mathcal{B}_Q$ is the set of indices corresponding to the Q-operators, 
$\mathbf{\Delta}(\tau-\tau^{\prime})$ is the hybridization
matrix and $H_0^{imp}$ contains one term, for example $i=0$, from the local
Hamiltonian (\ref{ham2}).  Using the expansion procedures described in the
previous section, with the hybridization replacing the ``hopping'' matrix $%
\mathbf{E}$, we obtain for the impurity Green's function a relation similar
to (\ref{lark2}),
\begin{equation}
\mathbf{G}_{imp}(i\omega_n) = \left[\mathbf{M}_{imp}^{-1}(i\omega_n) -
\mathbf{\Delta}(i\omega_n)\right]^{-1},
\end{equation}
where all the quantities are local, and $\mathbf{M}_{imp}$ is the
irreducible cumulant with respect to the hybridization. For practical
calculations, as well as to gain some physical insight, it is useful to
express the impurity problem in a Hamiltonian form. Taking into account that
in our matrix formulation the indices $(\alpha\beta)$ corresponding to
fermionic and bosonic Hubbard operators do not mix, all the matrices,
including $\Delta^{\alpha \beta ~ \alpha^{\prime}\beta^{\prime}}$, are
block-diagonal with one block corresponding to the fermionic indices and one
to the bosonic sector. Consequently, we will introduce two types of
auxiliary degrees of freedom, one associated with a fermionic bath,
described by operators $(a_{k\sigma},a_{k\sigma}^{\dagger})$, and one
associated with a bosonic bath, described by $(b_{k\sigma},b_{k\sigma}^{%
\dagger})$. The impurity Hamiltonian will be
\begin{eqnarray}
H_{imp} &=& H_0^{imp} + \sum_{(\alpha\beta)\in \mathcal{B}_Q} h_{imp}^{\alpha\beta} X^{\alpha\beta} +
\sum_{k,\sigma} \epsilon_k^a a_{k\sigma}^{\dagger}a_{k\sigma} +
\sum_{k,\sigma} \epsilon_k^b b_{k\sigma}^{\dagger}b_{k\sigma}  \nonumber \\
&+& \sum_{k}\sum_{(\alpha \beta)\in\mathcal{F}} [V_a^{\alpha \beta}(k)
~a_{k\sigma_{\alpha \beta}}^{\dagger} X^{\alpha \beta} + h.c.] +
\sum_{k}\sum_{(\alpha \beta)\in\mathcal{B}} [V_b^{\alpha \beta}(k)
~b_{k\sigma_{\alpha \beta}}^{\dagger} X^{\alpha \beta} + h.c.].  \label{Himp}
\end{eqnarray}
The first term in the right hand side describes an isolated impurity, the
second term represents the coupling to the external field and the summation is restricted to labels corresponding to bosonic operators having a non-zero mean value, the Q-operators ($\mathcal{B}_Q$), the third and
forth terms describe the fermionic and bosonic baths, respectively, and the
last two terms represent the coupling of the impurity with the two baths. In
the last two terms, the summations over the X-operator labels $(\alpha \beta)
$ are restricted to the fermionic ($\mathcal{F}$) and bosonic sectors ($%
\mathcal{B}$), respectively. The expression of the hybridization in terms of
coupling constants $V^{\alpha \beta}(k)$ can be obtained by integrating out
the auxiliary degrees of freedom. We have
\begin{equation}
\Delta^{\alpha\beta~\alpha^{\prime}\beta^{\prime}}(i\omega_n) = \frac{1}{N}%
\sum_k \frac{V_l^{\alpha \beta}(k)V_l^{\alpha^{\prime}\beta^{\prime}}(k)}{%
i\omega_n - \epsilon_k^{l}},
\end{equation}
where $l=a$ if the element is in the fermionic sector of the $\Delta$ matrix
and $l=b$ for the bosonic case.

Returning to our main problem, we  obtain the self-consistency conditions by
requiring that the lattice irreducible cumulant be equal to the effective
impurity problem irreducible cumulant, $\mathbf{M}_{imp}(z) = \mathbf{M}%
_{0}(z)$. By comparing the diagrammatic expansion of $\mathbf{M}_{imp}$ in
terms of bare cumulants (dressed with external field lines) and
hybridization functions with the expansion of $\mathbf{M}_{0}$, we observe
that the equality is satisfied if we identify a line representing $%
\Delta^{\mu\nu}$ with the sum of all the paths starting with $%
E_{0i}^{\mu\rho}$ and ending with $E_{j0}^{\pi\nu}$, as well as 
the sum betweeen an external field line $h^{\mu}$ and the 
one-particle reducible decoration 
$\sum_{\nu}\Delta^{\mu\nu} \langle X^{\nu} \rangle$
with the lattice one-particle reducible decoration $\sum_{j, \nu}
E_{ij}^{\mu~\nu} \langle X_j^{\nu} \rangle$.
Explicitly, the expansion for the impurity irreducible cumulant is identical
with the expansion of the lattice irreducible cumulant provided
\begin{equation}
\mathbf{\Delta}(i\omega_n) = \sum_{i,j}~ \mathbf{E}_{0i}\mathbf{G}%
_{ij}^{(0)}(i\omega_n)\mathbf{E}_{j0},  \label{self1}
\end{equation}
where $\mathbf{G}_{ij}^{(0)}(i\omega_n)$ represents the matrix Green's
function for a lattice with the site ``0'' removed, together with a
condition that determines self-consistently the external field $h$,
\begin{equation}
h_{imp}^{\mu} = \sum_{j, ~\nu\in \mathcal{B}_Q} E_{0j}^{\mu\nu} \langle
X_j^{\nu} \rangle - \sum_{\nu} \Delta^{\mu\nu}(0) \langle
X_0^{\nu} \rangle.  \label{selfh}
\end{equation}
Notice that equation (\ref{self1}) represents the generalization of the well
known DMFT self-consistency condition\cite{review}. In addition, if the
non-local Hamiltonian $H_1$ contains Hubbard operators that acquire a
non-zero mean value, our lattice model is mapped into an impurity model with
an external field determined self-consistently by equation (\ref{selfh}).

Our final step consists in finding a self-consistent equation that
determines the hybridization $\Delta(z)$ in terms of the lattice Green's
function. Comparing Eq. (\ref{lark1}) for $\mathbf{G}_{ij}$ (after making $%
\mathbf{M}_{ij} = \delta_{ij}\mathbf{M}_0$) with the equivalent expression
for $\mathbf{G}_{ij}^{(0)}$,
\begin{equation}
\mathbf{G}_{ij}^{(0)}(i\omega_n) = \delta_{ij}\mathbf{M}_0(i\omega_n) +
\sum_{k\neq 0}\mathbf{M}_0(i\omega_n)\mathbf{E}_{ik}\mathbf{G}%
_{kj}^{(0)}(i\omega_n)~~~~~~~~~ \mbox{for}~~i,j\neq0,
\end{equation}
we obtain for $i=j=0$:
\begin{eqnarray}
\mathbf{G}_{00}(i\omega_n) &=& \mathbf{M}_0(i\omega_n) + \mathbf{M}%
_0(i\omega_n)\sum_{k,l}~ \mathbf{E}_{0k}\mathbf{G}_{kl}^{(0)}(i\omega_n)%
\mathbf{E}_{l0}~\mathbf{M}_0(i\omega_n) + ~.~.~.  \nonumber \\
&=& \mathbf{M}_0\left[\mathbf{1} -\sum_{k,l}~ \mathbf{E}_{0k}\mathbf{G}%
_{kl}^{(0)}(i\omega_n)\mathbf{E}_{l0}~\mathbf{M}_0(i\omega_n)\right]^{-1}.
\end{eqnarray}
Using equation (\ref{self1}) for $\mathbf{\Delta}(i\omega_n)$ and the
Fourier transform for the local Green's function, we obtain the
self-consistent condition that determines the hybridization,
\begin{equation}
\frac{1}{N} \sum \left[\mathbf{M}_0^{-1}(i\omega_n) - \mathbf{E}(\mathbf{k})%
\right]^{-1}=\left[\mathbf{M}_0^{-1}(i\omega_n) - \mathbf{\Delta}(i\omega_n)%
\right]^{-1}.  \label{self0}
\end{equation}

Formally, Eq. (\ref{self0}) closely resembles the standard form of the DMFT
self-consistency condition\cite{review}. We should, however, keep in mind
the fact that this is a matrix equation involving Green's functions for the
Hubbard operators. To solve a problem described by the Hamiltonian (\ref%
{ham1}-\ref{ham3}) within this scheme, implies to solve the impurity
problem, (\ref{impur}), find an irreducible cumulant matrix $\mathbf{M}_0$,
then use the self-consistency condition (\ref{self0}) together with equation
(\ref{selfh}) to determine the new hybridization, $\mathbf{\Delta}(i\omega_n)
$, and the new field $\mathbf{h}_{imp}$. The process is repeated until convergence
is reached.

\subsection{The Hubbard Model}

In order to get a feeling of how this approach works, as well as to check
its validity, we can do a simple exercise by applying the procedure to
the Hubbard model, for which we should be able to recover the known DMFT
results. In this particular case, the Hamiltonian will be
\begin{equation}
H_{HUB} = \sum_{i}[-\mu(X_i^{\uparrow\uparrow} + X_i^{\downarrow\downarrow})
+(U-2\mu)X_i^{22}] + \sum_{i,j}\sum_{\sigma}~t_{ij}[X_i^{\sigma 0}X_j^{0
\sigma} + X_i^{2 \bar{\sigma}}X_j^{\bar{\sigma} 2} + \sigma X_i^{\sigma
0}X_j^{\bar{\sigma} 2} +\sigma X_i^{2 \bar{\sigma}}X_j^{0 \sigma}],
\label{hbham}
\end{equation}
where $\mu$ represents the chemical potential, $U$ the on-site Coulomb
interaction and $t_{ij}$ the hopping matrix elements, which are equal to $-t$
if i and j are nearest neighbors and zero otherwise. We can introduce the
hopping matrix,
\begin{equation}
\mathbf{E}(\mathbf{k}) = \left(%
\begin{array}{cc}
E^{\sigma 0 ~ 0 \sigma}(\mathbf{k}) & E^{\sigma 0 ~ \bar{\sigma} 2}(\mathbf{k%
}) \\
E^{2 \bar{\sigma} ~ 0 \sigma}(\mathbf{k}) & E^{2 \bar{\sigma} ~ \bar{\sigma}
2}(\mathbf{k}) \\
&
\end{array}%
\right) = \left(%
\begin{array}{cc}
\epsilon_k & \sigma \epsilon_k \\
\sigma \epsilon_k & \epsilon_k \\
&
\end{array}%
\right),
\end{equation}
where $\epsilon_k$ is the Fourier transform of $t_{ij}$. Next, we introduce
the Green's function matrix and the irreducible cumulant matrix:
\begin{equation}
\mathbf{G} = \left(%
\begin{array}{cc}
G^{0 \sigma ~ \sigma 0} & G^{0 \sigma ~ 2 \bar{\sigma}} \\
G^{\bar{\sigma} 2 ~ \sigma 0} & G^{\bar{\sigma} 2 ~ 2 \bar{\sigma}} \\
&
\end{array}%
\right); ~~~~~~~~~~ \mathbf{M} = \left(%
\begin{array}{cc}
M^{0 \sigma ~ \sigma 0} & M^{0 \sigma ~ 2 \bar{\sigma}} \\
M^{\bar{\sigma} 2 ~ \sigma 0} & M^{\bar{\sigma} 2 ~ 2 \bar{\sigma}} \\
&
\end{array}%
\right).
\end{equation}
Our goal is to show that the matrix equation (\ref{self0}) reduces in this
particular case to the well known DMFT self-consistency condition,
\begin{equation}
\mathcal{G}_0^{-1}(i\omega_n) = i\omega_n +\mu - \Delta_0(i\omega_n) = \left[ \frac{1}{N}\sum_{%
\mathbf{k}}\frac{1}{i\omega_n + \mu - \epsilon_k - \Sigma(i\omega_n)} \right]^{-1} + \Sigma(i\omega_n),
\end{equation}
where $\Sigma(i\omega_n)$ represents the self-energy of the single-particle electron Green's function and $\Delta_0(i\omega_n)$ the hybridization of the effective single-impurity problem.
The
electron single-particle Green's function, $\mathcal{G}$, is given by a sum
of all the elements of $\mathbf{G}$, as shown in Eq. (\ref{elgreen}).
Similarly, we define the quantity
\begin{equation}
\mathcal{M}(i\omega_n) = M^{0 \sigma ~ \sigma 0}(i\omega_n) + M^{\bar{\sigma}
2 ~ 2 \bar{\sigma}}(i\omega_n) + \sigma[M^{0 \sigma ~ 2 \bar{\sigma}%
}(i\omega_n) + M^{\bar{\sigma} 2 ~ \sigma 0}(i\omega_n)].
\end{equation}
It is straightforward to show that the matrix Green's function has the
structure:
\begin{equation}
\mathbf{G}(\mathbf{k}, i\omega_n) = [\mathbf{M}^{-1}(i\omega_n) - \mathbf{E}(%
\mathbf{k})]^{-1} = \frac{1}{1-\epsilon_k \mathcal{M}(i\omega_n)}\mathbf{M}%
(i\omega_n) + \delta \frac{\epsilon_k}{1-\epsilon_k \mathcal{M}(i\omega_n)}
\left(%
\begin{array}{cc}
-1 & \sigma \\
\sigma & -1 \\
&
\end{array}%
\right),  \label{struct}
\end{equation}
where $\delta = \det\{\mathbf{M}\}$. Using Eq. (\ref{struct}) and the
expression (\ref{elgreen}) for the electron Green's function, together with
the definition of $\mathcal{M}$, we have
\begin{equation}
\mathcal{G}(\mathbf{k}, i\omega_n) = \frac{1}{\mathcal{M}^{-1}(i\omega_n) -
\epsilon_k},
\end{equation}
which shows that $\mathcal{M}(i\omega_n)$ is the single particle irreducible
cumulant, $\mathcal{M}^{-1}(i\omega_n) = i\omega_n + \mu - \Sigma(i\omega_n)$%
, where $\Sigma(i\omega_n)$ is the single particle self-energy. On the other
hand, making the summation over $k$ in Eq. (\ref{struct}) and introducing
the result in the self-consistency relation (\ref{self0}) gives:
\begin{equation}
\mathbf{\Delta} = \Delta_0 \left(%
\begin{array}{cc}
1 & \sigma \\
\sigma & 1 \\
&
\end{array}%
\right),  \label{eldelta}
\end{equation}
with
\begin{equation}
\Delta_0 = \frac{b}{a \delta}, ~~~~~~~~~ a = \frac{1}{n}\sum_{\mathbf{k}}
\frac{1}{1-\epsilon_k \mathcal{M}}, ~~~~~~~~~ b=\frac{\delta(a-1)}{\mathcal{M%
}}.
\end{equation}
Taking into account the structure of $\mathbf{\Delta}$ given by Eq. (\ref%
{eldelta}), the impurity problem described by the statistical operator (\ref%
{impur}) reduces to an Anderson single impurity problem with a hybridization
$\Delta_0$. Finally, we have
\begin{equation}
\Delta_0(i\omega_n) = \mathcal{M}^{-1}(i\omega_n) - \left[ \frac{1}{N}\sum_{%
\mathbf{k}}\frac{1}{\mathcal{M}^{-1}(i\omega_n) - \epsilon_k} \right]^{-1},
\end{equation}
which represents the standard DMFT self-consistency equation.

%&&&&&&&&&&&&&&&&&&&&&&&&&&&&&&&&&&&&&&&&&&&&&&&&&&&&&&&&&&&&&&&&&

\section{A Functional Approach}

Functionals of Green's functions provide an elegant and powerful tool for
formulating dynamical mean-field theories. Within this framework, the DMFT
equations are obtained as a direct result of a simple approximation on the
Baym-Kadanoff functional\cite{review}. The extended dynamical mean-field
equations have also been rigorously derived using a functional technique\cite%
{edmft5}. Moreover, the functional perspective seems to offer the natural
framework for cluster generalizations of the dynamical mean-field theory
that include effects of short-range correlations. The purpose of this
section is to introduce a strong-coupling generalization of the functional
DMFT formulation\cite{shvaik}. The construction of a generalized Baym-Kadanoff-type
functional will allow us to present an alternative derivation of the
generalized DMFT equations, and, more importantly, will set the basis for a
future cluster extension of the G-DMFT approach.

The main idea of the functional interpretation of DMFT is to formulate a
local approximation for the Baym-Kadanoff functional $\Phi[G]$, which is the
sum of all vacuum-to-vacuum skeleton (two-particle irreducible) diagrams
constructed with the full propagator $G$ and the interaction vertices. This
functional has the property that
\begin{equation}
\Sigma_{ij}(i\omega_n) = \frac{\delta\Phi}{\delta G_{ji}(i\omega_n)}~~,
\label{phideriv}
\end{equation}
where $\Sigma_{ij}(i\omega_n)$ is the self-energy. As the coordination
number of the lattice goes to infinity, the self-energy becomes local and $%
\Phi$ will depend only on the local Green's functions, $G_{ii}$:
\begin{equation}
\Phi = \sum_{i}~\phi[G_{ii}],
\end{equation}
where $\phi[G_{ii}]$ is a functional of the local Green's function at site $i
$ only. This scheme cannot be transfered directly to approach our
strong-coupling problem. As we mentioned above, within our strong-coupling 
perspective, the natural quantity to describe the system
is the two-point irreducible cumulant, $M_{ij}(i\omega_n)$, rather than the self-energy. Our task is to
identify a conjugate quantity (which cannot be the Green's, function but
rather something with dimensions of energy), and to construct a functional
of this quantity that will determine $M$ by a relation similar to (\ref%
{phideriv}).

Following Refs. \onlinecite{shvaik} and \onlinecite{shvaik1}, we start by introducing two time dependent external sources, one , $%
J_{ij}^{\mu\nu}(\tau-\tau^{\prime})$, coupled to the X-operator Green's
functions, and the other, $\lambda_i^{\alpha\beta}(\tau)$, coupled to the average
$Q_i^{\alpha\beta} = \langle X_i^{\alpha\beta}\rangle$ of the bosonic Hubbard
operators that occur in the non-local Hamiltonian $H_1$. The corresponding
generating functional will be
\begin{equation}
\Omega[J,\lambda] = -\frac{1}{\beta} \ln Tr \hat{\rho}[J,\lambda],
\end{equation}
where the statistical operator $\rho$ is given by
\begin{equation}
\hat{\rho}[J,\lambda] = e^{-\beta H_0} \hat{T} \exp\left\{-\int_0^{\beta}d%
\tau\left[\int_0^{\beta}d\tau^{\prime}\left(H_1(\tau, \tau^{\prime}) +
\sum_{i,j}\sum_{\mu,\nu}J_{ji}^{\nu\mu}(\tau-\tau^{\prime})
X_i^{\mu}(\tau)X_j^{\nu}(\tau^{\prime})\right) + \sum_{i, \mu}
\lambda_i^{\mu}(\tau) X_i^{\mu}(\tau) \right] \right\}.
\end{equation}
The field $J_{ij}^{\mu\nu}(\tau-\tau^{\prime})$ is equivalent with
introducing frequency dependent coupling constants $E_{ji}^{\nu\mu}(i%
\omega_n) = E_{ji}^{\nu\mu} + J_{ji}^{\nu\mu}(i\omega_n)$. The Green's
functions can be expressed as functional derivatives of the generating
functional, $\Omega[J,\lambda]$, with respect to the frequency dependent
field,
\begin{equation}
G_{ij}^{\mu\nu}(i\omega_n) - \beta\langle X_i^{\mu} \rangle\langle X_j^{\nu}
\rangle = -\beta\frac{\delta \Omega[J,\lambda]}{\delta
E_{ji}^{\nu\mu}(i\omega_n)}\Bigg|_{J=0, \lambda=0},  \label{fderivG}
\end{equation}
where the average value $Q_i^{\mu} = \langle X_i^{\mu} \rangle$ is
non-zero only if $X_i^{\mu}$ is a Q-operator. 
The generating functional $\Omega[J,\lambda]$ reduces to
the grand-canonical potential, $\Omega$, in the absence of the external
sources.

The next step is to integrate Eq. (\ref{fderivG}). As we are interested in a
strong-coupling equivalent of the Baym-Kadanoff functional, we will express
the Green's function in terms of the two-point irreducible cumulant, rather
than the self-energy\cite{shvaik}. Using the notation $\hat{\mathbf{A}}$ for
a tensor with elements $A_{ij}^{\mu\nu}$, we can formally write equation (%
\ref{lark1}) as
\begin{equation}
\hat{\mathbf{G}} = \left[\hat{\mathbf{I}} - \hat{\mathbf{M}}\hat{\mathbf{E}}%
\right]^{-1}\hat{\mathbf{M}},  \label{hatG}
\end{equation}
where $\hat{\mathbf{I}}$ is the identity tensor, $I_{ij}^{\mu\nu} =
\delta_{ij}\delta_{\mu\nu}$. By integrating (\ref{fderivG}) with $\hat{%
\mathbf{G}}$ given by (\ref{hatG}) we obtain
\begin{equation}
\Omega[J,\lambda] = \Omega_0 + \frac{1}{\beta} \sum_{\omega_n}Tr~\ln[\hat{%
\mathbf{I}} - \hat{\mathbf{M}}\hat{\mathbf{E}}] + \frac{1}{\beta}
\sum_{\omega_n}Tr[\hat{\mathbf{M}}\hat{\mathbf{{\mathcal{E}}}}] + \Xi[%
J,\lambda],  \label{gama1}
\end{equation}
with $\Omega_0$ the grand-canonical potential in the atomic limit. The
quantity $\hat{\mathbf{{\mathcal{E}}}}$ is the renormalized coupling
constant (or renormalized generalized hopping) and can be represented
diagrammatically as a sum of chains of ``hopping'' lines and irreducible
cumulants. Analytically we have
\begin{equation}
{\mathcal{E}}_{ij}^{\mu\nu} = E_{ij}^{\mu\nu} +
\sum_{p,q}\sum_{\alpha,\beta}~ E_{iq}^{\mu\alpha} M_{qp}^{\alpha\beta}
E_{pj}^{\beta\nu} + .~.~. = \left[\hat{\mathbf{E}}(\hat{\mathbf{I}} - \hat{%
\mathbf{M}}\hat{\mathbf{E}})^{-1}\right]_{ij}^{\mu\nu}.  \label{calE}
\end{equation}
The last term in Eq. (\ref{gama1}) is a functional with the property
\begin{equation}
-\beta\frac{\delta \Xi[J,\lambda]}{\delta E_{ji}^{\nu\mu}(i\omega_n)} = Tr%
\left[\hat{\mathbf{M}}\frac{\delta \hat{\mathbf{{\mathcal{E}}}}}{\delta
E_{ji}^{\nu\mu}(i\omega_n)}\right] -\beta\langle X_i^{\mu} \rangle\langle
X_j^{\nu} \rangle.  \label{xideriv1}
\end{equation}
Also, because the functional derivative of $\Omega[J,\lambda]$ with respect
to $\lambda$ gives the average value of $X_i^{\mu}$, we will have
\begin{equation}
\frac{\delta\Xi}{\delta\lambda_i^{\mu}} = -\frac{1}{\beta}Tr\left[\hat{%
\mathbf{M}}\frac{\delta \hat{\mathbf{{\mathcal{E}}}}}{\delta
\lambda_i^{\mu}}\right] + \langle X_i^{\mu} \rangle.
\label{xideriv2}
\end{equation}

The standard way to proceed is to perform a Legendre transform of the
generating functional with respect to the interacting single-particle
Green's function and to eliminate the external source $J$ in favor of $G$ in
the functionals. Instead, by analyzing equations (\ref{xideriv1}) and (\ref%
{xideriv2}), we identify the renormalized coupling constant ${\mathcal{E}}%
_{ij}^{\mu\nu}$ as the quantity conjugate with the irreducible cumulant and
use it as an independent variable. Consequently, we introduce the new
functionals $\Omega^*[{\mathcal{E}}, \lambda; E]$ and $\Xi^*[{\mathcal{E}},
\lambda; E]$, with
\begin{equation}
\Omega^*[{\mathcal{E}}[J,\lambda], \lambda; E+J] = \Omega[J,\lambda],
~~~~~~~~~~~~ \Xi^*[{\mathcal{E}}[J,\lambda], \lambda; E+J] = \Xi[J,\lambda].
\end{equation}
The dependence on the last variable, $E$, involves only the 
components of the coupling constant tensor corresponding to Q-operator indices
and we have
\begin{equation}
\frac{\delta \Xi^*}{\delta E_{ij}^{\mu~\nu}} = \langle
X_i^{\mu}\rangle \langle X_j^{\nu}\rangle.  \label{derivXX}
\end{equation}
The functional derivative with respect to $\lambda$ of both $\Omega^*$ and $%
\Xi^*$ will give the average value of the bosonic X operator, while for the
Luttinger-Ward-type functional $\Xi^*$ we have
\begin{equation}
-\beta\frac{\delta \Xi^*}{\delta {\mathcal{E}}_{ji}^{\nu\mu}(i\omega_n)} =
M_{ij}^{\mu\nu}(i\omega_n).  \label{fderivM}
\end{equation}
This equation is the strong-coupling equivalent of (\ref{phideriv}) and
determines the irreducible two-point cumulant as a functional of the
renormalized ``hopping'' matrix, the bare diagonal coupling elements and the
external fields. The stationarity condition $\delta\Omega^* / \delta {%
\mathcal{E}}_{ji}^{\nu\mu} = 0$ yields the relation between the renormalized
couplings and the cumulants, ${\mathcal{E}}_{ij}^{\mu\nu} = [\hat{\mathbf{E}}%
(\hat{\mathbf{I}} - \hat{\mathbf{M}}\hat{\mathbf{E}})^{-1}]_{ij}^{\mu\nu}$,
which is identical to the definition (\ref{calE}). Again, at stationarity
and in the absence of the external fields the functional $\Omega^*$ reduces
to the grand canonical potential. 

For the case with no Q-operators in the non-local Hamiltonian our
construction is complete. In this particular case the only relevant
dependence in the functionals is on the renormalized coupling ${\mathcal{E}}$%
. The functional $\Omega^*[{\mathcal{E}}]$ is formally analogous to the
functional $\Gamma_{new}[G]$ introduced by Chitra and Kotliar\cite{edmft5}
from a weak-coupling perspective. In our picture the strong coupling
correspondent of the self-energy is, as noted before, the irreducible
cumulant, while the renormalized and bare coupling constants correspond to
the full and non-interacting Green's function, respectively. Also, the
strong-coupling analog of the Luttinger-Ward functional $\Phi[G]$ is $\Xi^*[{%
\mathcal{E}}]$. Similarly to its weak-coupling counterpart, $\Xi^*[{\mathcal{%
E}}]$ can be expressed as a sum of all vacuum-to-vacuum skeleton diagrams
containing bare cumulants and renormalized ``hopping'' lines. All the
skeleton diagrams are two-particle irreducible, i.e. they cannot be
separated into disconnected parts by cutting two ${\mathcal{E}}$ lines.

\begin{figure}[tbp]
\begin{center}
\includegraphics[width=0.5\textwidth]{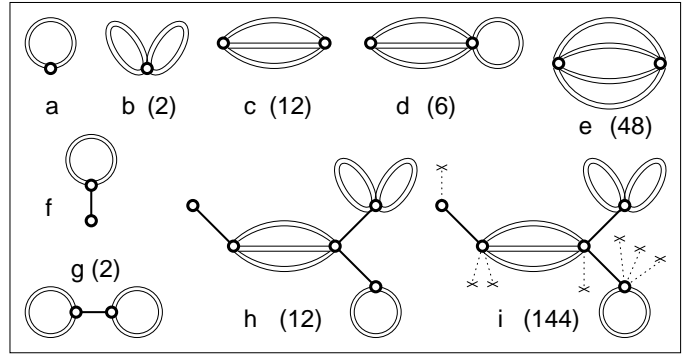}
\end{center}
\caption{Generalized skeleton diagrams contributing to the functional $\Xi^*$%
. The contribution of the external field $\protect\lambda$ is represented
only in diagram (i) by dotted lines. The numbers in parentheses represent
the symmetry factors (if different from 1) of the corresponding diagrams
(the number of ways in which a graph can be made isomorphic with itself).
The renormalized coupling ${\mathcal{E}}$ is represented by double lines.}
\label{FIG4}
\end{figure}

\begin{figure}[tbp]
\begin{center}
\includegraphics[width=0.4\textwidth]{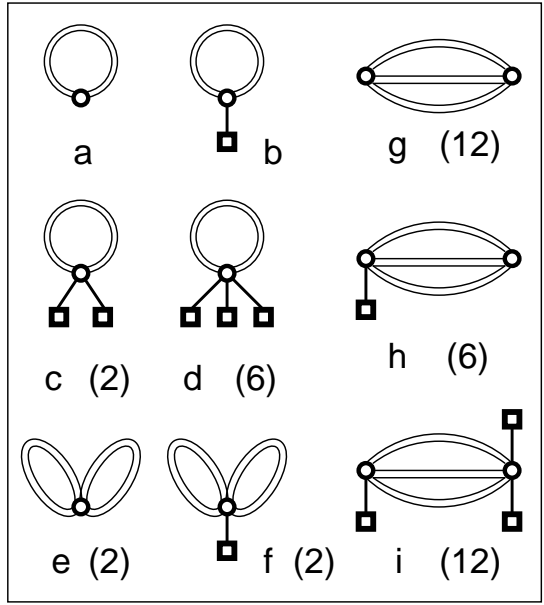}
\end{center}
\caption{Example of diagrams contributing to $\Psi[{\mathcal{E}},Q]$. The
average of a diagonal operator, $Q_i^{\protect\alpha}$, is represented by a
square. The symmetry factors are written in parentheses. If $Q_{i}^{\protect%
\mu}=0$, the remaining basic
skeleton diagrams (a, e, g, ...) are identical with the remaining diagrams
in Fig. \protect\ref{FIG4} (a, b, c, ...), and $\Psi = \Xi^*$. In general,
we have the basic diagrams plus diagrams obtained by decorating the bare
cumulants with lines and squares representing $E_{ij}^{\protect\mu\protect\nu} Q_j^{\protect\nu}$.}
\label{FIG5}
\end{figure}

In general, however, the diagrammatic expression of $\Xi^*$ is more
complicated and involves, in addition to the bare cumulants, both
renormalized coupling constants, ${\mathcal{E}}_{ji}^{\nu\mu}$, and bare
 coupling constants, $E_{ij}^{\mu\nu}$. Also, lines
corresponding to the external field $\lambda$ have to be added. The
generalized skeleton diagrams have to satisfy two conditions: i) Each
diagram is two-particle irreducible with respect to the renormalized
``hopping''; ii) By cutting any bare ``hopping'' line the diagram is divided
into two disconnected parts. These conditions determine a tree-like
structure for the generalized skeleton diagrams with skeleton blocks coupled
by bare $E$ lines. This structure is illustrated in Fig. \ref{FIG4} for a
few low-order contributions to $\Xi^*$. Due to the tree-like structure of
the diagrams, it is rather difficult to find a local approximation for the
functional $\Xi^*$. In order to overcome this problem, we proceed to the last
step of our construction and define the Legendre transform of $\Omega ^*$
with respect to the average of bosonic Hubbard operators, $Q_i^{\mu} =
\langle X_{ij}^{\mu}\rangle$,
\begin{equation} 
\Gamma[{\mathcal{E}}, Q] = \Omega^*[{\mathcal{E}}, \lambda; E] -  
\sum_i\sum_{\mu\in \mathcal{B}_Q} \lambda_i^{\mu}Q_i^{\mu},  \label{gamaeq1} 
\end{equation}
where the bare coupling $E_{ij}^{\mu\nu}$ is treated as a   
parameter and is not written explicitly as an argument of the functional $\Gamma 
$. Using the condition $-\lambda_i^{\mu} = \delta\Gamma / \delta 
Q_i^{\mu}$ to eliminate $\lambda$, we obtain 
\begin{equation}
\Gamma[{\mathcal{E}}, Q] = \Omega_0 + \frac{1}{\beta} \sum_{\omega_n}Tr~\ln[{%  
\hat{\mathbf{I}} - \hat{\mathbf{M}}[{\mathcal{E}}, Q]\hat{\mathbf{E}}}] + 
\frac{1}{\beta} \sum_{\omega_n}Tr[\hat{\mathbf{M}}[{\mathcal{E}}, Q]\hat{%  
\mathbf{{\mathcal{E}}}}] - \frac{1}{2}\sum_{i,j}\sum_{\mu,\nu\in \mathcal{B}_Q} Q_i^{\mu} 
E_{ij}^{\mu\nu} Q_j^{\nu} + \Psi[{\mathcal{E}}, Q].  \label{gama2} 
\end{equation}
The properties with respect to ${\mathcal{E}}$ of the new functionals are 
similar to those of $\Omega^*$ and $\Xi^*$, namely the stationarity 
condition $\delta\Gamma / \delta {\mathcal{E}}_{ji}^{\nu\mu} = 0$ yields 
equation (\ref{calE}), and we have 
\begin{equation}  
-\beta\frac{\delta \Psi[{\mathcal{E}},Q]}{\delta {\mathcal{E}}% 
_{ji}^{\nu\mu}(i\omega_n)} = M_{ij}^{\mu\nu}(i\omega_n).  \label{fderivM1}
\end{equation}
In addition, stationarity with respect to $Q$, $\delta\Gamma / \delta 
Q_i^{\mu} = 0$, yields 
\begin{equation} 
\frac{\delta \Psi[{\mathcal{E}},Q]}{\delta Q_i^{\mu}} = \sum_{j, ~\beta\in \mathcal{B}_Q}
E_{ij}^{\mu\nu} ~Q_j^{\nu}.  \label{fderivQ} 
\end{equation}
The diagrammatic representation of the new Luttinger-Ward-type functional $%
\Psi$ contains skeleton diagrams that are two-particle irreducible with
respect to the ${\mathcal{E}}$ lines and have cumulants dressed with $EQ$ 
contributions. The structure of the diagrams is illustrated in Fig \ref{FIG5}% 
. Using the diagrammatic representation of the functionals $\Psi$ and $\Xi^*$% 
, as well as that of $Q$, and taking into account the symmetry factors 
associated with each diagram, it is straightforward to check that 
\begin{equation} 
\Psi[{\mathcal{E}},Q] = \Xi^*[{\mathcal{E}},\lambda=0; E] + \frac{1}{2}% 
\sum_{i,j}\sum_{\mu,\nu\in \mathcal{B}_Q} Q_i^{\mu} E_{ij}^{\mu\nu} Q_j^{\nu}, 
\end{equation} 
where $Q_i^{\mu} = \langle X_i^{\mu} \rangle = \delta\Xi^* / \delta
\lambda_i^{\mu} |_{\lambda = 0}$.

We are prepared now to introduce the generalized DMFT from a functional
perspective. By analogy with the standard approach\cite{review}, we
construct an approximate theory by making the ansatz that the Luttinger-Ward
functional $\Psi$ depends only on local renormalized ``hopping'' matrices,
i.e.,
\begin{equation}
\Psi[{\mathcal{E}},Q] = \sum_i \psi[\mathbf{{\mathcal{E}}}_{ii},Q],
\end{equation}
where $\psi$ is a functional that depends only on the local renormalized
coupling constants at site i. In diagrammatic terms, it means that all the
bare cumulants contained in a skeleton diagram contributing to $\psi$ are
defined on the same site. A direct consequence of this ansatz is that the
two-point irreducible cumulant of the theory is local,
\begin{equation}
\mathbf{M}_{ij}(i\omega_n) = \frac{\delta\Psi}{\delta \mathbf{{\mathcal{E}}}%
_{ii}(i\omega_n)}\delta_{ij},  \label{Mdr}
\end{equation}
with $\delta_{ij}$ being the Kronecker $\delta$ symbol. The stationarity
condition for $\Gamma$ yields
\begin{equation}
\mathbf{{\mathcal{E}}}_{ii} = \left[\hat{\mathbf{E}}(\hat{\mathbf{I}} - \hat{%
\mathbf{M}}\hat{\mathbf{E}})^{-1}\right]_{ii} = \frac{1}{N}\sum_k \mathbf{E}%
(k)(\mathbf{I} - \mathbf{M}\mathbf{E}(k))^{-1},  \label{calEloc}
\end{equation}
where $M$ is given by (\ref{Mdr}) and, for the second equality, we assumed
that the system is homogeneous. Relation (\ref{calEloc}) represents our
generalized DMFT equation and it should be viewed as a functional equation
for the local renormalized coupling matrix. To cast it in a more familiar
form, let us observe that using the expressions for the Green's function,
Eq. (\ref{hatG}), and for the renormalized coupling, Eq. (\ref{calE}), we
have
\begin{equation}
\hat{\mathbf{G}} = \hat{\mathbf{M}}\hat{\mathbf{{\mathcal{E}}}}\hat{\mathbf{M%
}} + \hat{\mathbf{M}}.
\end{equation}
Taking into account the local character of the irreducible cumulant, we
obtain, for a homogeneous system,
\begin{equation}
\mathbf{G}_{ii}(i\omega_n) = \frac{1}{N}\sum_k \left[\left(\frac{\delta\Psi}{%
\delta \mathbf{{\mathcal{E}}}_{ii}}\right)^{-1} - \mathbf{E}(k)\right]^{-1},
\label{calEloc1}
\end{equation}
where $\mathbf{G}_{ii}(i\omega_n)$ has to be understood as a functional of ${%
\mathcal{E}}$.

To be able to solve Eq. (\ref{calEloc}) or Eq. (\ref{calEloc1}) we need to 
know the functional dependence of $\Psi$, at least approximately. A direct 
approach is rather hard in practice, therefore it is more convenient to 
generate the functionals from a purely local theory. A simple inspection of 
the diagrammatic expression for $\Psi$ (see Figure \ref{FIG5}), shows that,
in the case when all the bare cumulants are defined on the same site $i=0$,
the same diagrams will correspond to an impurity problem defined by the
statistical operator given by Eq. (\ref{impur}). In fact, taking into 
account the previous construction of the strong-coupling functionals, we
have
\begin{eqnarray}
\Gamma_{imp}[{\mathcal{D}}, Q] = \frac{\Omega_0}{N} 
&+& \frac{1}{\beta}
\sum_{\omega_n}Tr~\ln[{\mathbf{I} - \mathbf{M}_{imp}[{\mathcal{D}}, Q]\mathbf{%
\Delta}}] + \frac{1}{\beta} \sum_{\omega_n}Tr[\mathbf{M}_{imp}[{\mathcal{D}}% 
, Q]\mathbf{{\mathcal{D}}}]  \\ \nonumber 
&-& \frac{1}{2}\sum_{\mu, \nu\in \mathcal{B}_Q} Q^{\mu} \Delta^{\mu\nu}(0)Q^{\nu}  -\sum_{\mu \in \mathcal{B}_Q} h_{imp}^{\mu}Q^{\mu}  + \Psi_{imp}[{\mathcal{D}}, Q],  \label{gimp} 
\end{eqnarray} 
where $\mathbf{{\mathcal{D}}} = \mathbf{\Delta}[\mathbf{I} - \mathbf{M}_{imp}\mathbf{\Delta}]^{-1}$ is the renormalized hybridization,   
$Q^{\mu} = \langle X^{\mu}\rangle$ and $\Delta^{\mu\nu}(0)$ is the  
hybridization at zero frequency. In addition, we have to keep in mind that the bare impurity cumulants are dressed by external field lines, $h_{imp}^{\mu}$.  The diagrammatic expressions
for $\Psi[{\mathcal{E}}, Q]$ and $\Psi_{imp}[{\mathcal{D}}, Q]$ are
identical and we have $\Psi / N = \Psi_{imp}$ provided 
\begin{equation}
{\mathcal{E}}_{ii}^{\mu\nu} = {\mathcal{D}}^{\mu\nu}; ~~~~~~~~~~~~~~~~
\sum_{j, \nu} E_{0j}^{\mu\nu}Q_j^{\nu} = h_{imp}^{\mu} + \sum_{\nu}\Delta^{\mu\nu}(0)Q^{\nu}.  \label{scsce} 
\end{equation}
Equations (\ref%
{scsce}) represent the self-consistency condition for our strong-coupling generalized
dynamical mean-field theory. The first of these equations can be rewritten
in the more familiar form, Eq. (\ref{self0}), by simply using the
expressions of the renormalized quantities ${\mathcal{E}}$ and ${\mathcal{D}}
$. A direct consequence of the first self-consistency condition is the
equality of the irreducible cumulants, $\mathbf{M}_{ii}(i\omega_n) = \mathbf{%
M}_{imp}(i\omega_n)$. Finally, let us mention that this correspondence with
the impurity problem is also useful in establishing a relation between the
grand-canonical potential of the lattice and the grand-canonical potential
of the impurity. At stationarity both $\Gamma_{imp}$ and $\Gamma$ reduce
to the corresponding grand-canonical potential, while $\Psi / N = \Psi_{imp}$.
Therefore we have
\begin{equation}
\frac{\Omega}{N} = \Omega_{imp}
+\frac{1}{2} \sum_{\mu,\nu\in \mathcal{B}_Q} \left[ Q^{\mu} \Delta^{\mu\nu}(0)Q^{\nu}
 - \frac{1}{N} \sum_{i,j}
Q_i^{\mu} E_{ij}^{\mu\nu} Q_j^{\nu}\right] - \frac{1}{\beta}\sum_{\omega_n}\left[Tr \ln[\mathbf{I} - \mathbf{M}\mathbf{\Delta}]  -
\frac{1}{N} \sum_k Tr \ln [\mathbf{I} - \mathbf{M}\mathbf{E}(k)] \right].
\end{equation}

\begin{acknowledgments}
This work was supported by the NSF under grant  DMR-0096462 and by the Rutgers University Center for Materials Theory. We would like to thank C. Henley and A. Schiller for useful discussions and S. Pankov and A. Shvaika for helpful critical observations.    
\end{acknowledgments}
%&&&&&&&&&&&&&&&&&&&&&&&&&&&&&&&&&&&&&&&&&&&&&&&&&&&&&&&&&&&&&&&&&

\appendix

\section{Alternative derivation of the G-DMFT equations}

The purpose of this Appendix is twofold: 1) To 
present an alternative formulation of our
strong-coupling theory that is closely related to the more familiar
perturbation expansions based on Wick's theorem and translate the basic 
quantities involved in this approach in the corresponding language. 
2) To propose a scheme that allows one to expand around the DMFT solution.  
The central idea is to
decouple the nonlocal term using a generalized Hubbard-Stratonovich
transformation and to trade in the X-operators for a set of canonical
fermionic and bosonic fields. In terms of these new auxiliary fields, the
standard techniques of the perturbation theory can be applied. This
formulation represents a generalization of the strong-coupling expansion
used by Pairault, S\'en\'echal, and Tremblay\cite{tramb} for the study of
the Hubbard model.

Starting with the partition function
\begin{equation}
Z = Tr\left\{e^{\beta H_0} ~\hat{T} \exp\left[-\int_0^{\beta} d\tau
\sum_{<i,j>} \sum_{\mu, \nu} X_i^{\mu}(\tau) E_{ij}^{\mu\nu} X_j^{\nu}(\tau)%
\right]\right\},  \label{partfunc}
\end{equation}
we perform a Hubbard-Stratonovich transformation and express the non-local
part as a Gaussian integral over the auxiliary field $\zeta_i^{\mu}$,
\begin{equation}
Z = Z_0\int\mathcal{D}\zeta~\exp\left\{\int_0^{\beta} d\tau \sum_{<i,j>}
\sum_{\mu, \nu}\zeta_i^{\mu}(\tau) (\chi_0^{-1})_{ij}^{\mu\nu}
\zeta_j^{\nu}(\tau) - S_{int}[\zeta]\right\},  \label{zdezeta}
\end{equation}
with
\begin{equation}
S_{int}[\zeta] = - \ln\Bigg\langle \hat{T} \exp\left\{-\int_0^{\beta} d\tau
\sum_{i, \mu} \left[\zeta_i^{\mu}(\tau)X_i^{\mu}(\tau) +
X_i^{\mu}(\tau)\zeta_i^{\mu}(\tau)\right]\right\}\Bigg\rangle_0.
\label{Sdezeta} 
\end{equation}
The field $\zeta_i^{\mu}$ contains both fermionic (Grassmann variables) and
bosonic (complex) components, as determined by the corresponding
X-operators. The free propagator appearing in the quadratic term of Eq. (\ref%
{zdezeta}) is $\chi_{0~ij}^{\mu\nu} = E_{ij}^{\mu\nu}$. Notice that the
auxiliary fields have canonical fermionic or bosonic statistics and all the
complications related to the unconventional algebra of the Hubbard operators
are buried in the average $\langle .~.~.\rangle_0$ with respect to the
unperturbed ensemble. Due to the nature of the local Hamiltonian $H_0$, the
interaction term $S_{int}[\zeta]$ can be written as a sum of purely local
contributions, each having an expression similar to Eq. (\ref{Sdezeta}) but
without the summation over the site index $i$. Expanding in the $\zeta$%
-fields, we have explicitly
\begin{equation}
S_{int}[\zeta] = \sum_i\sum_{n=0}^{\infty} \frac{1}{n!}\sum_{\mu_1, ...,
\mu_n} C_i^n(\mu_1,..., \mu_n;\tau_1, ...,\tau_n)~
\zeta_i^{\mu_n}(\tau_n)...\zeta_i^{\mu_1}(\tau_1),  \label{intvert}
\end{equation}
where $C_i^n$ is the n-point bare cumulant defined by Eq. (\ref{cumul0}). We
can now develop a standard perturbation theory for the $\zeta$-field, having
$\chi_0 = E$ as the free propagator, and vertices given by $C_i^n$. The
relation between the Green's function of the X-operators and that of the
auxiliary fields can be easily obtained if we notice that, due to the form
of the interaction term (\ref{Sdezeta}), we have
\begin{equation}
G_{\mu\nu}(i,j;\tau,\tau^{\prime}) = \frac{1}{Z} \int\mathcal{D}\zeta~\frac{%
\delta^2 S_{int}[\zeta]}{\delta \zeta_j^{\nu}(\tau^{\prime})\delta
\zeta_i^{\mu}(\tau)} \exp\{-S_0[\zeta]\},
\end{equation}
where $S_0[\zeta]$ represents the quadratic term in Eq. (\ref{zdezeta}).
After performing two integrations by part, we obtain
\begin{equation}
\hat{\mathbf{G}} = -\hat{\mathbf{\chi}}_0^{-1} + \hat{\mathbf{\chi}}_0^{-1}%
\hat{\mathbf{\chi}}\hat{\mathbf{\chi}}_0^{-1},
\end{equation} 
where we used the tensor notation, and we introduced the full propagator, $% 
\chi$, for the auxiliary fields, defined as
\begin{equation}
\chi_{ij}^{\mu\nu}(\tau, \tau\prime) = -\langle \hat{\mbox{T}} \{\zeta_i^{\mu}(\tau)\zeta_j^{\nu}(\tau')\}\rangle + \langle \zeta_i^{\mu}\rangle\langle \zeta_j^{\nu}\rangle,
\end{equation} 
where $\hat{\mbox{T}}$ is the (imaginary) time ordering operator and $\langle .~.~.\rangle$ represents the average over the statistical ensemble.
 In terms of the self-energy for the $\zeta$ 
variables, $\Sigma_{\zeta} = \chi_0^{-1} - \chi^{-1}$, one has  
\begin{equation} 
\hat{\mathbf{G}} = (\hat{\mathbf{\Sigma}}_{\zeta}^{-1} - \hat{\mathbf{\chi}}%
_0)^{-1},
\end{equation}
which shows that $\Sigma_{\zeta}$ corresponds to the irreducible cumulant $M$
of the X-operator formulation. Also, the full Green's function $\hat{\mathbf{%
\chi}} = (\hat{\mathbf{\chi}}_0^{-1} - \hat{\mathbf{\Sigma}}_{\zeta})^{-1} =
\hat{\mathbf{E}}(\hat{\mathbf{I}} - \hat{\mathbf{M}}\hat{\mathbf{E}})^{-1}$
is the renormalized coupling costant $\hat{\mathbf{\mathcal{E}}}$ of the
original theory. In addition, in the presence of bosonic Hubbard operators 
with non-zero mean value, we have 
\begin{equation}
\langle \zeta_i^{\mu}\rangle = \sum_{j, \nu} 
E_{ij}^{\mu\nu} \langle X_j^{\nu}\rangle,
\end{equation}
showing that the average of the auxiliary field represents the effective 
\textquotedblleft magnetic\textquotedblright\  field of the original 
formulation. 
This equation completes the correspondence between the the formulation in
terms of X-operators and that in terms of auxiliary $\zeta$-fields.

It is now straightforward to re-derive our generalized mean-field equations
 using the standard techniques 
of a perturbation theory for the auxiliary fields. For example, the 
Baym-Kadanoff functional, $\Phi[\chi]$, can be defined in the usual way 
as the sum of all vacuum-to-vacuum skeleton diagrams. The corresponding   
free-energy functional can be written as 
\begin{equation} 
\Lambda[\chi, Q] = \frac{1}{\beta}\sum_{\omega_n} Tr\ln\left[\hat{\mathbf{\chi}}%
_0^{-1} - \hat{\mathbf{\Sigma}}_{\zeta}[\chi,Q]\right] + \frac{1}{\beta}% 
\sum_{\omega_n} Tr[\hat{\mathbf{\Sigma}}_{\zeta}[\chi,Q] \hat{\mathbf{\chi}}]
-\frac{1}{2}\sum_{i,j}\sum_{\mu,\nu}Q_i^{\mu}(\chi_0)_{ij}^{\mu\nu}Q_j^{\nu} + \Phi[\chi, Q],  \label{fren} 
\end{equation} 
where the possibility of a non-zero average, $Q$, for the X-operators has been
explicitly  considered. 
Taking into account the correspondence between the auxiliary field   
formulation and the original X-operator formulation, we can see that $\Lambda  
[\chi, Q]$ is the equivalent of the functional $\Gamma[\mathcal{E}, Q]$, while 
the Baym-Kadanoff functional $\Phi[\chi, Q]$ corresponds to $\Psi[\mathcal{E}, Q]$, up to 
an $\mathcal{E}$-independent term equal to the atomic grand-canonical    
potential $\Omega_0$. 

\bigskip 

We can construct now a strong-coupling expansion to a certain 
order in $(E_{ij}/U)$, as in the work of Pairault, S\'en\'echal, and Tremblay%
\cite{tramb}. However, any finite order expansion will intrinsically contain 
the problem of the exponentially large degeneracy of the atomic ground  
state, leading  to the break-down of the solution in the low temperature  
limit\cite{tramb}. 
Next, we can recover our strong coupling DMFT scheme as a local approximation 
for the  Baym-Kadanoff functional. This  renormalized perturbation approach 
avoids the problem of the low temperature barrier.  
Further, we can exploit the idea of introducing auxiliary degrees of freedom
and construct a formulation which, in principle, allows one to determine
corrections to the DMFT approximation. The basic idea is to introduce the
auxiliary fields in such a way as to obtain the DMFT result as the lowest
order approximation. Starting again from the partition function (\ref%
{partfunc}), instead of performing directly a Hubbard-Stratonovich
transformation, we first add and subtract in the exponent a term equal to
\begin{equation}
\int_0^{\beta} d\tau \int_0^{\beta} d\tau^{\prime}\sum_{i} \sum_{\mu, \nu}
X_i^{\mu}(\tau) \Delta_{ii}^{\mu\nu}(\tau-\tau^{\prime})
X_i^{\nu}(\tau^{\prime}),
\end{equation}
where $\Delta_{ii}^{\mu\nu}(\tau-\tau^{\prime})$ is the hybridization
function for the impurity problem associated with the DMFT approximation.
Notice that this is a purely local quantity. Next, we introduce the
auxiliary field $\xi$ that decouples the quadratic term containing $%
\chi_0(\tau - \tau^{\prime}) = E\delta(\tau-\tau^{\prime}) -
\Delta(\tau-\tau^{\prime})$. The effective theory for the new variables will
have a free term
\begin{equation}
S_0[\xi] = - \int_0^{\beta} d\tau \int_0^{\beta} d\tau^{\prime}\sum_{<i,j>}
\sum_{\mu, \nu} \xi_i^{\mu}(\tau)
(\chi_0^{-1})_{ij}^{\mu\nu}(\tau-\tau^{\prime}) \xi_j^{\nu}(\tau^{\prime}),
\end{equation}
and an interaction term
\begin{equation}
S_{int}[\xi] = -\ln\Bigg\langle \hat{T} \exp\left\{ -\int_0^{\beta} d\tau 
d\tau^{\prime}~X_i^{\mu}\Delta_{ii}^{\mu\nu} X_i^{\nu} + \xi_i^{\mu}X_i^{\mu} 
+ X_i^{\mu}\xi_i^{\mu}\right\}\Bigg\rangle_0,  \label{sintdmft} 
\end{equation} 
where, for simplicity, we omitted the summations and the imaginary time 
dependence. Again, the interaction part can be written as a sum of purely 
local terms. However, the vertices of the new theory are no longer the bare 
cumulants of the Hubbard operators. Instead, let us notice that each term $% 
S_{int}[\xi_i]$ of the interaction part represents the generating  
functional $-\ln[Tr(\{\hat{\rho}_{imp}\}]$ for the impurity problem  
defined by Eq. (\ref{impur}).  
In the present formulation, the external fields $h_{imp}^{\alpha\beta}$ are 
represented by the average values $\langle \xi^{\alpha\beta} \rangle$. A
direct consequence of this correspondence is that the second order cumulant
generated by $S_{int}$ is equal to the impurity Green's function $G_{imp}$
associated with the generalized DMFT solution of our problem. Further, the
relation between the propagator for the $\xi$-field and the X-operator
Green's function can be derived as before and we have
\begin{equation}
\hat{\mathbf{G}} = (\hat{\mathbf{\Sigma}}_{\xi}^{-1} - \hat{\mathbf{\chi_0}}%
)^{-1},  \label{Gdechi}
\end{equation}
with $\Sigma_{\xi}$ the self-energy for the $\xi$-field.

Let us construct now the lowest order approximations for the pertubation
expansions formulated in terms of auxiliaryy fields $\zeta$ and $\xi$,
respectively. In the first case, the simplest diagram contributing to $%
\Sigma_{\zeta}$ is equal to the atomic Green's function. The corresponding
result represents the Hubbard-I approximation\cite{tramb}. In the second
case, the zero-order approximation for the self-energy is given by the DMFT
impurity Green's function, $\Sigma_{\xi}^{(0)} = G_{imp}$. Using the
correspondence relation (\ref{Gdechi}) with $\chi_0 = E - \Delta$, as well
as the equation $G_{imp}^{-1}(i\omega_n) = i\omega_n - \Delta(i\omega_n) -
\Sigma_{imp}(i\omega_n)$, we have
\begin{equation}
\mathbf{G}(i\omega_n) = (\mathbf{G}_{imp}^{-1}(i\omega_n) - \mathbf{E} +
\mathbf{\Delta})^{-1} = [i\omega_n \mathbf{I} - \mathbf{E} - \mathbf{\Sigma}%
_{imp}(i\omega_n)]^{-1}.
\end{equation}
This shows that, in the formulation using auxiliary $\xi$ fields, the
generalized DMFT result is recoverd as the lowest order approximation.
Obviously, in order to determine the hybridization function $\Delta$, one
has to first find the solution of the problem within the DMFT scheme.
However, the strong coupling expansion constructed here allows us to
determine in a systematic way the corrections to the DMFT result by
computing higher order contributions to the self-energy $\Sigma_{\xi}$.

\end{document}